\documentclass[prb,twocolumn,english,showpacs,amsmath]{revtex4-1}
\usepackage{graphicx}
\usepackage{amssymb}
\usepackage{hyperref}

\begin{document}

\title{Controlling Chemical Reactions of a Single Particle}

\author{Lothar Ratschbacher, Christoph Zipkes, Carlo Sias, and Michael K{\"o}hl}

\affiliation{Cavendish Laboratory, University of Cambridge, JJ Thomson Avenue, Cambridge CB3 0HE, United Kingdom}

\maketitle
\textbf{The control of chemical reactions is a recurring theme in physics and chemistry. Traditionally, chemical reactions have been investigated by tuning thermodynamic parameters, such as temperature or pressure. More recently, physical methods such as laser\cite{Jones2006} or magnetic field\cite{Chin2010} control have emerged to provide completely new experimental possibilities, in particular in the realm of cold collisions. The control of reaction pathways is also a critical component to implement molecular quantum information processing\cite{Demille2002}. For these undertakings, single particles provide a clean and well-controlled experimental system. Here, we report on the experimental tuning of the exchange reaction rates of a single trapped ion with ultracold neutral atoms by exerting control over both their quantum states. We observe the influence of the hyperfine interaction on chemical reaction rates and branching ratios, and monitor the kinematics of the reaction products. These investigations advance chemistry with single trapped particles towards achieving quantum-limited control of chemical reactions and indicate limits for buffer gas cooling of single ion clocks.}

The full control over all quantum mechanical degrees of freedom of a chemical reaction allows for the identification of fundamental interaction processes and for steering chemical reactions. This task is often complicated in heteronuclear systems by a multitude of possible reaction channels, which make theoretical treatments very challenging. Therefore, focussing on the best-controlled experimental conditions, such as using state-selected single particles and low temperatures, is crucial for the investigation of chemical processes at the most elementary level. The hybrid system of trapped atoms and ions offers key advantages in this undertaking. On the one hand, ion traps offer a large potential well depth in order to trap the reaction products for precision manipulation and investigation. On the other hand, contrary to pure ionic systems, there is no Coulomb-barrier between the particles which fundamentally prevents chemical reactions at low temperatures. Therefore, the  efforts to control the motional degrees of freedom of one\cite{Willitsch2008,Staanum2008,Roth2006} and both\cite{Grier2009, Zipkes2010, Zipkes2010b, Schmid2010,Hall2011, Rellergert2011,Sullivan2012} reactants in hybrid atom-ion systems have paved new ways towards cold chemistry. The yet missing component is the simultaneous control of the internal degrees of freedom.

The interaction between an ion and a neutral atom at long distances is dominated by the attractive polarization interaction potential, which is of the form
\begin{equation}
  V(r) = -\frac{C_4}{2 r^4}.
	\label{eqn:dxs_C4_potential}
\end{equation}
Here, $C_4=\alpha_0 q^2/(4 \pi \epsilon_0)^2$ is proportional to the neutral particle polarizability $\alpha_0$, $q$ is the charge of the ion, $\epsilon_0$ is the vacuum permittivity, and $r$ is the internuclear separation. Inelastic collisions take place at short internuclear distances. In the cold, semiclassical regime this requires collision energies above the centrifugal barrier\cite{Langevin1905,Vogt1954,Makarov2003}. Such processes are referred to as Langevin type collisions and happen, even for cold collisions\cite{Grier2009,Zipkes2010b}, at an energy-independent rate $\gamma_{Langevin} = 2\pi \sqrt{C_4/\mu}\,n_a$. Here $\mu$ is the reduced mass of the collision partners and $n_a$ is the neutral atom density. More subtle effects, such as the hyperfine interaction, which may lead to atom-ion Feshbach resonances, are not included in the polarization potential and have been investigated theoretically only so far\cite{Idziaszek2009}.

Experimentally, reactive Langevin collisions in the polarization potential have been investigated in ground state collisions\cite{Grier2009, Zipkes2010, Zipkes2010b, Schmid2010,Rellergert2011,Ravi2011}, which have displayed relatively low rates for inelastic collisions, except for resonant charge exchange\cite{Grier2009}. Recently, first steps towards understanding reactive collisions in excited electronic states have been made using large ion crystals\cite{Hall2011,Rellergert2011,Sullivan2012}, suggesting either a dominant contribution from very short-lived electronic states in the Rb+Ca$^+$ system\cite{Hall2011} or, contrarily, a negligible contribution from excited state collisions in the Ca+Yb$^+$ system\cite{Rellergert2011}. 

Here, we demonstrate how control over the internal electronic state of a single ion and the hyperfine state of neutral atoms can be employed to tune cold exchange reaction processes. 
We study quenching, charge-exchange, and branching ratios, and we use near-resonant laser light to control the rates. Our measurements show a large sensitivity of the charge-exchange reaction rates to the atomic hyperfine state, highlighting the influence of the nuclear spin on atom-ion collisions. 

\begin{figure}
    \includegraphics[width=1.0\columnwidth]{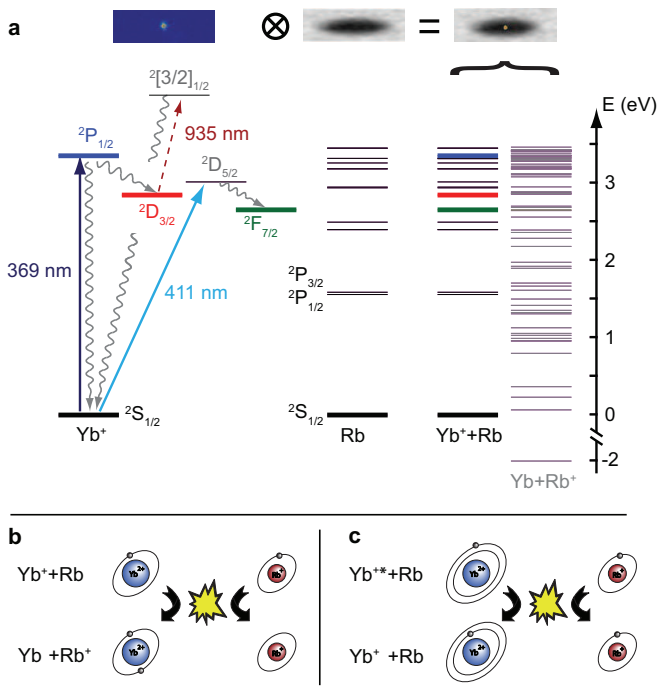}
	\caption[Level Scheme]{Level scheme and inelastic processes of the ion-neutral system. \textbf{a)} Left: Yb$^+$ level scheme with the transitions used for optically pumping the ion (to scale). Middle: Level scheme of a Rubidium atom. Right: Asymptotic level schemes of the two channels Yb$^+$+Rb and Yb+Rb$^+$. The collision is initiated in the Yb$^+$+Rb manifold and the Yb+Rb$^+$ manifold can be populated by charge-exchange processes. Pictorial representation of the charge exchange reaction \textbf{(b)} and quenching from an excited state \textbf{(c)}. Shown are the filled core electronic shells and the relevant valence electrons.}\label{fig1}
\end{figure}

In our experiment we study collisions between ultracold $^{87}$Rb atoms and single $^{174}$Yb$^+$ ions, for which $\gamma_{Langevin}/n_a =2.1\times10^{-15}$\,m$^3$/s. We start by determining the inelastic collision loss rate coefficients for the long-lived $^{2}D_{3/2}$ (radiative lifetime 52\,ms) and $^{2}F_{7/2}$ (radiative lifetime 10\, years) states of the ion (see Figure 1 and Methods). The collisional stability of these states is of importance in buffer gas cooled ion clocks\cite{Schauer2009} and quantum information processing. By optical pumping, both states are prepared as ``dark'' states (see Methods), in order to study pure two-body collisions without the presence of light. This approach differs fundamentally from previous experiments in atom-ion\cite{Hall2011} excited state collisions, which always have been in presence of near-resonant laser light. We measure the inelastic loss rate $\gamma_\ell$ by immersing the ion for a variable time $t$ into the neutral atom cloud and determining the survival probability\cite{Zipkes2010,Zipkes2010b,Ueberholz2000,Weber2010} $P_s = \exp\left(-\gamma_\ell t\right)$. In order to understand and model our data, we make three assumptions: Firstly, the atom and the ion can undergo an inelastic process only in a Langevin-type collision. Secondly, the characteristic collision time is significantly shorter than the radiative lifetime. Thirdly, inelastic collisions can only be exothermic thanks to the very low kinetic energy of the colliding partners and the absence of resonant laser light. We model  $\gamma_\ell=\epsilon\,\gamma_{Langevin}$ as proportional to the Langevin collision rate using the state-dependent proportionality constant $\epsilon$. For the electronic ground-state of the ion interacting with the $|F=2,m_F=2\rangle$ hyperfine ground state of Rb we find\cite{Zipkes2010b} $\epsilon_S^{|2,2\rangle}=\times 10^{-5\pm0.3}$, and for the excited states we measure $\epsilon_D^{|2,2\rangle}=1.0\pm 0.2$ and $\epsilon_F^{|2,2\rangle}=0.018\pm 0.004$ (see Table \ref{table:branchingRatios}). The constant $\epsilon=1$ corresponds to the largest allowed inelastic collision rate in the semiclassical model, and even in near-resonant charge-exchange between equal elements of atoms and ions\cite{Grier2009} it was expected and approximately found to be $\epsilon=1/2$. In this regard, our results are unexpected since the $^2D_{3/2}$ state of Yb$^+$ combined with the $^2S_{1/2}$-state of Rb is off-resonant by 0.14\,eV to the next available asymptotic state in the Yb+Rb$^+$ manifold (see Fig. 1). Quite oppositely, the $^2F_{7/2}$-state has nearby asymptotic states but its inelastic collision rate is significantly lower. 

We compare these results now to inelastic collisions in the absolute lowest hyperfine state $|F=1,m_F=1\rangle$ of the neutral atom. The hyperfine energy difference between the $|F=1,m_F=1\rangle$ and the $|F=2,m_F=2\rangle$ states of Rb is 30\,$\mu$eV, which is larger than the collision energy but negligible on the scale of the molecular potentials or the trap depth (250\,meV). We find a significantly enhanced probability for inelastic collisions with the ion in the electronic ground state $\epsilon_S^{|1,1\rangle}=(35\pm11)\times\epsilon_S^{|2,2\rangle}$, which demonstrates an important role of the hyperfine interaction. For collisions with the ion in the $D_{3/2}$ state, the value of $\epsilon$ remains unchanged.

For collisions in the electronic ground state of the ion, exothermic inelastic collisions are inevitably associated with a chemical reaction, leading to charge exchange or possibly the radiative association of molecules. The excess energy is converted into photons and/or kinetic energy of the reaction products. Depending on the amount of kinetic energy released, the reaction products are ejected from the ion trap or kept and detected by in-trap mass spectrometry\cite{Drewsen2004,Zipkes2010b}. A partial conversion of internal energy into kinetic energy in the predominantly predicted\cite{Makarov2003} case of radiative charge exchange could be caused by the energy-dependent Franck-Condon overlap of the wave function of the entrance channels (A$^1\Sigma^+$ and a$^3\Sigma^+$) and the exit channel $X^1\Sigma^+$ of the (RbYb)$^+$ potentials, owing to a 70$\%$ smaller polarizability of the $X^1\Sigma^+$ state (assuming negligible momentum transfer by the photon and constant orbital angular momentum). We control the distribution of the kinetic energy release in the charge exchange reaction by employing the hyperfine state of the neutral atoms; for atoms in $|F=1,m_F=1\rangle$ we retain Rb$^+$ with $(48\pm3)\%$  probability in the ion trap and for atoms in $|F=2,m_F=2\rangle$ we find it with $(35\pm3)\%$ probability. We also study inelastic collisions in the excited dark $D-$ and $F-$states, where the probabilities for observing Rb$^+$ are quite similar to each other (see Table \ref{table:branchingRatios}). We do not observe the formation of (YbRb)$^+$ or Rb$_2^+$ molecules. The search for the latter was performed with the isotope $^{176}$Yb$^+$ in order to achieve the required mass resolution.

\begin{table*}
\caption[Branching Ratios]{Measured proportionality constant $\epsilon$ and branching ratios.  The left part of the table refers to neutral atoms in the $|F=2,m_F=2\rangle$ state and the right part to neutral atoms in the $|F=1,m_F=1\rangle$ state. Loss events of the charged particle are always chemical reactions for the $S-$states, whereas for $D-$ and $F-$ states quenching and chemical reactions can contribute.}
\begin{tabular}{|l|l|l|l|l||l|l|}
\hline
 	                                    &  	$^{2}S_{1/2}$ & $^{2}D_{3/2}$ & $^{2}F_{7/2}$ & $^{2}P_{1/2}$ &  $^{2}S_{1/2}$ & $^{2}D_{3/2}$ \\
 	\hline
     $\epsilon$                         & $10^{-5\pm0.3}$ & $1.0\pm0.2$ & $0.018\pm0.004$ & $0.1\pm0.2$ &  $(35\pm11)\epsilon_S^{|2,2\rangle}$ & $1.0\pm0.2$\\

 	charged particle lost				 & $65\%$ 	& $87\%$  & $84\%$&	&  $50\%$ &\\
 	Rb$^{+}$ identified  & $35\%$  	& $12\%$	& $15\%$  & &  $48\%$ &\\
 	dark Yb$^{+}$ identified &	& $<1\%$ 	& 		&		& $<1\%$ &\\
 	hot ion (unidentified)		&	&      		& $1\%$  &	& $2\%$ &\\
  	number of events		& $283$	&   $754$   		&  $225$ & 	& $236$ &\\
 	\hline
\end{tabular}
\label{table:branchingRatios}
\end{table*}

\begin{figure}
    \includegraphics[width=.8\columnwidth]{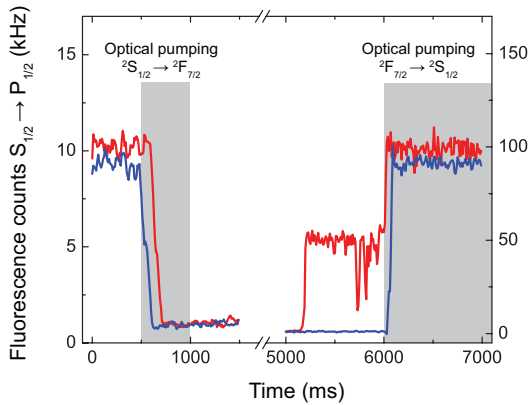}
	\caption[Short Name]{\label{fig:fStateSeq}Collisional quenching from $^{2}F_{7/2}$ to $^{2}S_{1/2}$. A two-ion Coulomb crystal is prepared in the $^2F_{7/2}$ dark state by optical pumping, then interacts with the neutral atoms, and is subsequently probed by laser fluorescence (see Methods). An early appearance of laser fluorescence indicates collisional quenching (red curve) as compared to no quenching (blue curve). The large fluorescence dips indicate a high temperature of the ion crystal.}
\end{figure}

With regards to collisional quenching between different electronic levels of the ion, we are principally able to detect two different quenching scenarios: $^{2}F_{7/2}\rightarrow\, ^{2}S_{1/2}$ and $^{2}D_{3/2} \rightarrow\, ^{2}F_{7/2}$. We directly observe quenching of $^{2}F_{7/2}\rightarrow\, ^{2}S_{1/2}$ as shown in Figure 2, where we exemplarily show two different experimental runs, with and without quenching (see also Methods). After the quenching, we observe an increased temperature of the ions, which indicates a release of kinetic energy smaller than the depth of our trap, despite of the large energy gap between the $^{2}S_{1/2}$ and the $^{2}F_{7/2}$ states. The quenching rate from $^{2}D_{3/2} \rightarrow\, ^{2}F_{7/2}$ was observed not to be detectable above our background rate.

\begin{figure}
    \includegraphics[width=1.0\columnwidth]{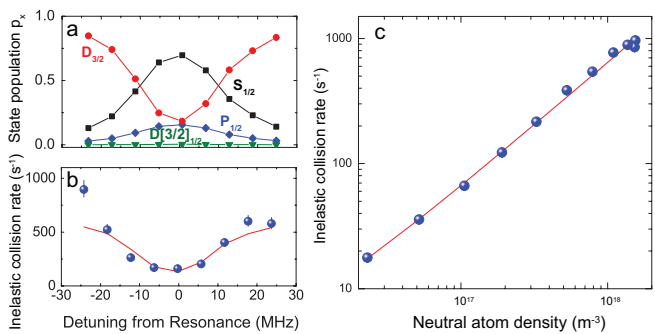}
	\caption[Short Name]{\label{fig:pdModel}Inelastic collision control by laser light. \textbf{a} State population measurement of the ion in the absence of neutral atoms as a function of the detuning of the repump laser at 935\,nm. \textbf{b} Inelastic collision rate for the same detuning data in presence of neutral atoms at $n_a=1\times 10^{18}$\,m$^{-3}$. The solid line shows the theoretical result of Eq.~\ref{eqn:etaFull}. \textbf{c} Density dependence of the inelastic collision rate with the repump laser on resonance. The exponent of the power-law fit (solid line) is $0.98\pm 0.02$. }
\end{figure}

Having established the inelastic collision parameters of the metastable $D$- and $F$-states without resonant laser light, we now turn our attention to inelastic collisions in the presence of laser light on both the $^2S_{1/2}-\,^2P_{1/2}$ (369 nm) and the $^2D_{3/2}-\,^3D[3/2]_{1/2}$ (935 nm) transitions. The purpose of the light is to experimentally tune the rates and to observe the occurrence of inelastic collisions in real time. The radiative lifetime of the $^2P_{1/2}$ state (8\,ns) is too short compared to the collision rate to provide pure P-state measurements. Here and in the following, we assume that the $F$-state is unoccupied since the 411-nm light is turned off and collisional quenching into the $F$-state has been measured to be negligible. Therefore, the overall inelastic collision rate is determined by a mixture of $S$-, $P$-, $D$-, and $D[3/2]$-states
\begin{equation}
	\gamma_\ell =2 \pi \sqrt{C_4/\mu}\,n_a \left(p_{S}\,\epsilon_{S} + p_{P}\,\epsilon_{P} + p_{D}\,\epsilon_{D}+p_{D[3/2]}\,\epsilon_{D[3/2]}\right).
	\label{eqn:etaFull}
\end{equation}
Here, $p_{x}$ is the occupation probability of state $x$, which we determine experimentally for different settings of laser intensities and detunings (see Methods). In Figure 3a we show the state populations of the $S$, $P$, $D$ and $^3D[3/2]_{1/2}$ states as we vary the frequency of the laser at 935\,nm on the $^2D_{3/2} -\, ^3D[3/2]_{1/2}$ transition. Figure 3b shows the associated change in the reaction rates in the presence of the neutral atoms, which are proportional to the inelastic collision rates using the branching ratios of Table 1. We demonstrate tuning of the reaction rate by one order of magnitude and we find it closely follows the model of equation (2) indicating a dominant contribution from the $D$-state. From these data we also extract $\epsilon_P=0.1\pm 0.2$, which is small and consistent with zero. Figure 3c shows the linear scaling of the inelastic collision rate with neutral atom density, confirming the picture of binary collisions.

\begin{figure}
    \includegraphics[width=1.0\columnwidth]{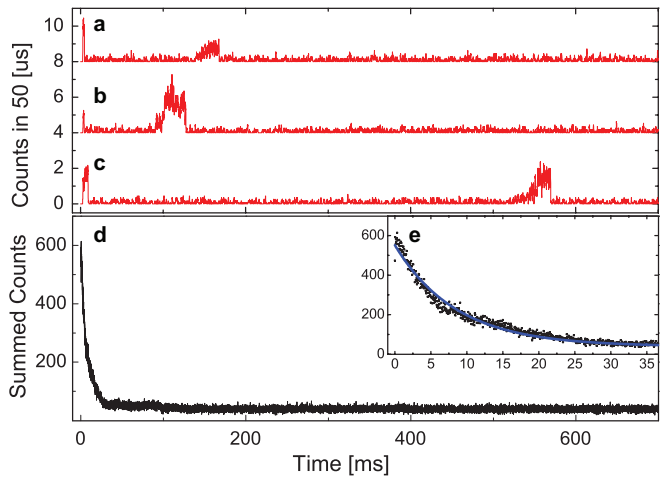}
	\caption[Short Name]{\label{fig:dynamics}Monitoring of inelastic atom-ion collisions. \textbf{a-c} Recorded fluorescence for selected events. After a quick initial loss, the fluorescence reoccurs after random times when the kinetic energy released in the inelastic collision has been removed by cooling. Then, a second collision occurs and the ion disappears again. The curves are vertically offset for clarity.  \textbf{d} Sum of 343 repetitions of the experiment, which is fitted with an exponential decay for short times. \textbf{e} Zoom-in on the initial decay of plot \textbf{d}. The solid line is an exponential fit to the data.}
\end{figure}

Finally, we turn our attention to the observation of the kinematics of the collision products. In Figure 4 we show the fluorescence at 369\,nm during the interaction, with the 935\,-nm light on. Subplots 4a-4c show individual experimental runs, in which the initial sharp loss of fluorescence results from an inelastic collision. If some kinetic energy is released in this process, the ion is on a large trajectory in the trap (much larger than the size of the atom cloud) but  can be re-cooled, which is signalled by the relatively slow increase of the reoccurring fluorescence from which we determine a lower bound for the release energy of 8\,meV. We observe in $4\%$ of our events that the fluorescence of the ion reoccurs after a certain time, before it undergoes a second inelastic collision and is dark again. As we keep the Yb$^+$ ion in the trap, these processes are not charge-exchange reactions but quenching processes with a kinetic energy release less than $\approx 250$\,meV. Since the ion is colliding in highly excited electronic states, such as $P_{1/2}$ and $D_{3/2}$ with internal energies of $\approx 3\,$eV, this suggests a mostly radiative decay into the ground state $S_{1/2}$. We have ruled out that the reoccurrence events are linked to the dissociation of a potential molecular ion in a secondary collision with neutral atoms\cite{Rellergert2011}. To this end, we have performed this measurement also with extremely short interaction times between neutral atoms and the ion (on the order of a few collision times) and performed mass spectrometry on the dark ion after the collision confirming the absence of a molecular ion.

We thank D. Sigle for experimental assistance, {EPSRC} (EP/H005676/1), {ERC} (Grant number 240335), and the Leverhulme Trust ({C.S.}) for support.

The experimental protocols were devised by all authors, data taking and data analysis was performed by L.R., C.Z., and C.S., and the manuscript was written by M.K. with contributions from all authors.

Correspondence and requests for materials should be addressed to C.S.~(email: cs540@cam.ac.uk).

\section*{methods}

\subsection{Preparation of ultracold atoms and ions}
We prepare $4\times 10^5$ neutral $^{87}$Rb atoms in the $|F=2,m_F=2\rangle$ hyperfine state of the electronic ground state at temperatures down to $T\approx 200$\,nK in a harmonic magnetic trap of characteristic frequencies\cite{Palzer2009} $(\omega_x,\omega_y,\omega_z)=2 \pi \times (8,26,27)$\,Hz. By changing atom number and temperature, we tune the atomic density by two orders of magnitude. The atoms can be transferred into an optical dipole trap formed by two crossed laser beams at 1064\,nm. Here, the atoms can be transferred into the ground hyperfine state $|F=1,m_F=1\rangle$ using a resonant microwave pulse. At the same location, we trap single Yb$^+$ ions in a radio-frequency Paul trap with secular trap frequencies of $\omega_\perp=2 \pi \times 150\,$kHz radially and $\omega_{ax}=2 \pi \times 42$\,kHz axially\cite{Zipkes2010,Zipkes2010b}. We use standard techniques to cool and detect single ions or small ion crystals on the cycling transition $^{2}S_{1/2}-\,^{2}P_{1/2}$ near 369\,nm wavelength (see Fig. 1). From the excited $^2P_{1/2}$ state, radiative decay populates the $^{2}D_{3/2}$ state with a probability\cite{Olmschenk2007} of $\approx1/200$. For efficient laser cooling and detection, a laser at 935\,nm pumps the population via the $^{3}D[3/2]_{1/2}$ state back into the cooling cycle. Preparation in $^{2}D_{3/2}$ (radiative lifetime 52\,ms) is achieved by optical pumping from the $S_{1/2}$ state using laser light at 369\,nm in less than $10\,\mu$s in the absence of light at 935\,nm, which is significantly faster than the inverse of the Langevin collision rate of a few $10^3$\,s$^{-1}$. Alternatively, the $^{2}F_{7/2}$ state (radiative lifetime 10\, years) is populated by optical pumping on the $^{2}S_{1/2} -\, ^{2}D_{5/2}$ line with spontaneous decay on the $^{2}D_{5/2} -\, ^{2}F_{7/2}$ transition\cite{Taylor1997}. The ion is spatially overlapped with the center of the ultracold neutral atom cloud by displacing the two independent trapping potentials. The typical collision energies\cite{Zipkes2011}, due to residual micromotion, are on the order of $\sim 50$\,mK.

\subsection{Detection of collisional quenching}
We prepare two ions in a small Coulomb crystal at time $t=0$ in the $^{2}S_{1/2}$ ground state, which we detect by near-resonant laser fluorescence at 369\,nm (together with repumping at 935\,nm). The optical pumping into the $^{2}F_{7/2}$ state using light at 411\,nm is switched on at $t=500$\,ms for 500\,ms and typically within 100\,ms the ion is pumped into the desired state (see Figure 2). As a result, the fluorescence counts on the 369-nm\, transition drop to zero. After the interaction with the neutral atoms (effective duration typically 16\,ms) and the subsequent removal of the neutral atom cloud, the fluorescence on the $^{2}S_{1/2}-\,^{2}P_{1/2}$ transition (together with repumping at 935\,nm) is probed from $t=5000$\,ms onwards. If the ion is still in the $^{2}F_{7/2}$ state it will not scatter photons, however, if it has been quenched to $^{2}S_{1/2}$, we observe fluorescence. At $t=6000$\,ms, the ions are optically pumped from the $F$-state back into the $S$-state to ensure that no ions have been lost. The fluorescence count rate during the preparation part of the sequence is lower owing to the presence of an offset field from the magnetic atom trap, which is turned off when the atoms are released.

\subsection{Experimental determination of the ion electronic state occupation}
Owing to the experimental situation of a multilevel system in presence of a magnetic field and imperfect polarization of the laser beams, we determine the state populations $p_x=\tau_x/\tau_c$ experimentally rather than relying on theoretical modelling. Here, $\tau_c$ is the average time between two spontaneous decays on the $^2P_{1/2}-\,^2D_{3/2}$ transition, $\tau_x$ is the average time spent in state $x\epsilon \{S,P,D,D[3/2]\}$ during $\tau_c$, and $\tau_c=\sum_x \tau_x$. The time $\tau_P$ is given by the decay rate $\Gamma$ of the $P_{1/2}$ state and the branching ratio of the $P$-state into $S-$ and $D-$states as $\tau_P=200/\Gamma$. A sequence of alternating pulses of the 369\,nm and the 935\,nm lasers are used to determine $\tau_{S}$ and $\tau_{D}$. $\tau_{P}+\tau_{S}$ is observed as the exponential decay constant of fluorescence after the 369 nm laser is pulsed on. The pulse length is set to get complete depletion of the $S$-$P$ system into the $D$-state. For a given setting of intensity and detuning of the 935-nm laser, the average number of photons $N_{369}$ per 369-nm pulse depends on the duration $t_{935}$ of the preceding repump pulse. Varying the length $t_{935}$ allows us to retrieve $\tau_{D}$ and the photon detection efficiency $\eta$ from the fit $200 \eta[1-\exp(-t_{935}/\tau_{D})]=N_{369}$ to the counted photon number. We find $\eta=(2.1\pm0.1)\times 10^{-3}$, in accordance with the numerical aperture of the imaging system and the quantum efficiency of the single photon counter. The lifetime of the excited $^3D[3/2]_{1/2}$ state is $\tau_{D[3/2]}=40$\,ns and the branching ratio is $98\%$ into $S_{1/2}$ and $2\%$ into $D_{3/2}$. Decay into $P_{1/2}$ is not dipole-allowed due to parity\cite{Biemont1998} and therefore we do not have to consider a cascade through intermediate levels.

\end{document}